\documentclass{tPHM2e}

\usepackage{epstopdf}

\begin{document}

\jvol{96} \jnum{20} \jyear{2016} \jmonth{June}


\title{X-Ray diffraction on large single crystals using a powder diffractometer}

\author{
\name{A. Jesche\textsuperscript{a,b}$^{\ast}$\thanks{$^\ast$Corresponding author. Email: anton.jesche@physik.uni-augsburg.de}, 
M. Fix\textsuperscript{a}, 
A. Kreyssig\textsuperscript{b,c}, 
W. R. Meier\textsuperscript{b,c} and 
P. C. Canfield\textsuperscript{b,c}
}
\affil{
\textsuperscript{a}Center for Electronic Correlations and Magnetism, Institute of Physics, University of Augsburg, D-86159 Augsburg, Germany;\\
\textsuperscript{b}The Ames Laboratory, Iowa State University, Ames, Iowa, USA\\
\textsuperscript{c}Department of Physics and Astronomy, Iowa State University, Ames, USA
}
\received{published 17 Jun 2016}
}

\maketitle

\begin{abstract}
Information on the lattice parameter of single crystals with known crystallographic structure allows for estimations of sample quality and composition.
In many cases it is sufficient to determine one lattice parameter or the lattice spacing along a certain, high-symmetry direction, e.g. in order to determine the composition in a substitution series by taking advantage of Vegard's rule. 
Here we present a guide to accurate measurements of single crystals with dimensions ranging from 200\,$\mu$m up to several millimeter using a standard powder diffractometer in Bragg-Brentano geometry.
The correction of the error introduced by the sample height and the optimization of the alignment are discussed in detail.
In particular for single crystals with a plate-like habit, the described procedure allows for measurement of the lattice spacings normal to the plates with high accuracy on a timescale of minutes.
\end{abstract}


\section{Introduction}
X-Ray powder diffraction is one of the most suitable and direct methods for phase analysis and the determination of lattice parameters. 
The normal procedure requires a grinding of the starting material, which could be single crystals or a conglomerate of polycrystalline material, to a fine powder. 
Ideally, a flat specimen of randomly oriented crystallites is prepared and placed in the center of the diffractometer circle (Bragg-Brentano geometry).
However, there are several cases where the grinding of samples is not possible or not suitable:
For example, the material could be too malleable and 'smears out' instead of breaking into small crystallites\,\cite{Ni2008b,Hodovanets2014,Ran2014} or the available amount is not sufficient for a powder sample (e.g. LiIr single crystals\,\cite{Jesche2014c}). 
Furthermore, it is sometimes advantageous to select a particular single crystal out of a larger batch (e.g. for measuring the specific heat, magnetization, etc.) \textit{based} on the lattice parameter, followed by further measurements of the physical properties.
Not of the least concern is the crime of grinding a beautiful, well-faceted single crystal into a barren powder. 

However, in many cases it is possible to perform X-Ray diffraction measurements on single crystals in full analogy to powder measurements, e.g. Refs.\,\cite{Mahan1990, Zaumseil1994, Hanzig2013}. 
The measurements can be carried out on compact, commercially available diffractometers in a fast and easy to adopt experimental routine.
Including sample mounting the whole procedure can be done within a couple of minutes without any recalibration or hardware adjustment of the diffractometer.
In particular for plate-like single crystals with large surfaces perpendicular to high-symmetry directions the procedure is straightforward.
Such plate-like morphologies are known for tetragonal, hexagonal, orthorhombic and even monoclinic crystals.
When the movement of sample holder ($\omega$) and X-Ray detector (2$\theta$) are coupled ($\theta/2\theta$), then only lattice planes parallel to the surface of the sample holder are accessible (the same situation occurs in $\theta$\,-\,$\theta$ geometry for diffractometer with $\theta_{\rm incident}$ and $\theta_{\rm outgoing}$ as mechanical axis). 
For simplicity and since $\omega$ and 2$\theta$ are indeed mechanically coupled in several commercially available diffractometers we restrict our considerations to this case (the extension to variable $\omega$/$2\theta$ or $\theta$\,-\,$\theta$ geometries is straightforward).

This paper is meant to provide a guide to accurate measurements of single crystals using a powder diffractometer that does not require vast experience in X-Ray diffraction.
In section\,\ref{hohe} the correction of the error introduced by the sample height is discussed in order to eliminate the main error source in such measurements. 
Other errors (such as the use of a flat rather than a curved specimen or the sample transparency) are neglected. 
The measurements performed on standard materials indicate a relative error for the lattice parameter of $\sim 0.1$\,\%, provided the diffractometer is well aligned (which was checked by corresponding powder diffraction measurements). 
In section\,\ref{tilt} we address the measurement of samples that are small and/or show a significant misalignment.
A brief discussion and summary is given in section\,\ref{ende}.

\section{Sample height correction}\label{hohe}

\begin{figure}
\begin{center}
\includegraphics[width=0.85\textwidth]{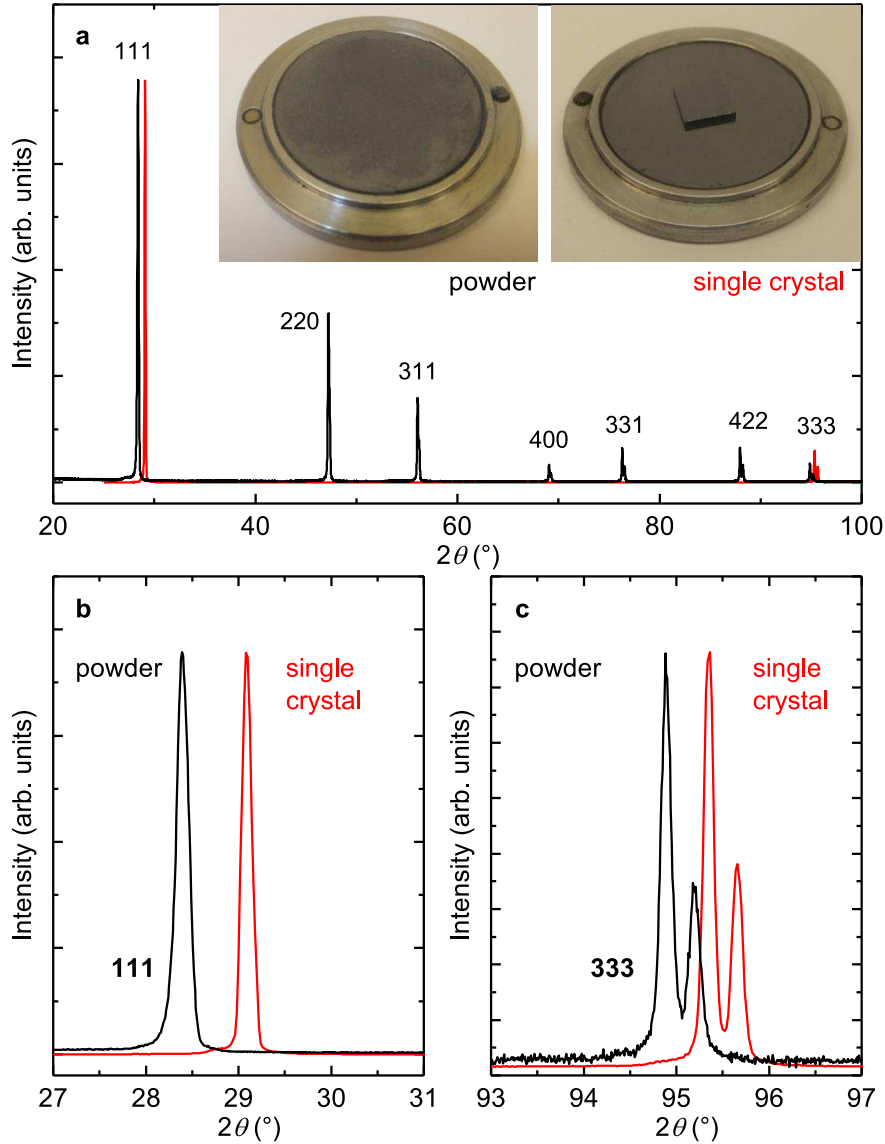}
\caption{X-Ray diffraction pattern of silicon measured on polycrystalline powder and on a single crystal in (111) orientation (Cu-$K\alpha$ radiation). a) Due to the sample height (0.78 mm) the two accessible Bragg reflections of the single crystal ($1\,1\,1$ and $3\,3\,3$) are significantly shifted to higher angles. Insets: silicon powder and a single crystal on zero-background sample holders as used for the measurement. The enlarged views shown in the panels b) and c) reveal a shift of $0.68^\circ$ for the $1\,1\,1$ reflection that is reduced to $0.46^\circ$ for the $3\,3\,3$ reflection. The $K\alpha_{1,2}$ splitting of the $3\,3\,3$ reflection is clearly visible in both samples.}
\label{silicon}
\end{center}
\end{figure}
The surface of a powder specimen can be adjusted to coincide with the diffractometer axis with a high degree of accuracy. 
On the other hand, mounting a single crystal at the correct height and simultaneously keeping the surface parallel to the sample holder can be an arduous task. 
However, the effect of a sample displacement on the Bragg peak position is crucial (see e.g. Ref.\,\cite{tablesc}).
This is shown in Fig.\,\ref{silicon} for the diffraction pattern of silicon measured on a powder and on a single crystal, respectively. 
The measurements were performed using a Rigaku Miniflex\,2 diffractometer with Cu-$K\alpha_{1,2}$ radiation (graphite monochromator). 
The intensity was normalized with respect to the $1\,1\,1$ peak.
The silicon powder was sprinkled on a zero-background sample holder. 
The displacement of the powder surface is close to zero as confirmed by a LeBail fit and the small value of the obtained sample height (see below). 
The single crystal, a small piece of a wafer, is oriented along (1\,1\,1) and, accordingly, only two reflections are available for $2\theta = 20^\circ-100^\circ$ (those are $1\,1\,1$ and $3\,3\,3$). 
The single crystal has a thickness of 0.78\,mm, a typical size for samples investigated in basic solid state research. This value corresponds to the theoretical displacement. 
Both, the powder sample and the single crystal, as mounted for the measurement, are shown as insets in Fig.\,\ref{silicon}.
 
Enlarged views of the diffraction patterns around the $1\,1\,1$ and the $3\,3\,3$ Bragg peaks are given in Figs.\,\ref{silicon}b and c, respectively.
The peak positions for Cu-$K\alpha_{1}$ wavelength was determined from a Pseudo-Voigt fit. 
For the powder, the $1\,1\,1$ Bragg peak appears at $2\theta = 28.39^\circ$ in reasonable agreement with the literature data (NIST standard 640d, $2\theta = 28.44^\circ$).
For the single crystal, however, the peak position is shifted by $\Delta2\theta = 0.68^\circ$ to 29.07$^\circ$.
The difference in the peak positions is somewhat smaller for the $3\,3\,3$ Bragg peak: $\Delta2\theta = 0.46^\circ$ (Fig.\,\ref{silicon}c, intensities are normalized). 
The peak position of the powder, $2\theta_{3 3 3} = 94.89^\circ$, is again close to the expected value of $2\theta = 94.95^\circ$, whereas a significantly larger value of $2\theta_{333} = 95.35^\circ$ is obtained for the single crystal.
The good agreement of the powder data with the literature values indicates that the deviation observed for the single crystal is dominated by the sample height ($\equiv$ displacement). 
Other error sources, like a $2\theta$-offset of the detector, misalignment, or transparency are negligible in this case.

\begin{figure}
\begin{center}
\includegraphics[width=0.57\textwidth]{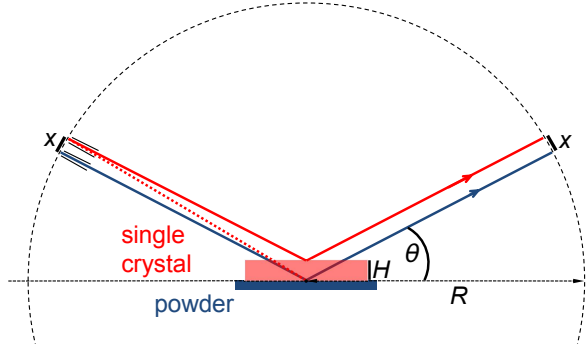}
\caption{Schematic of the scattering in Bragg-Brentano geometry. The X-Ray path for an ideally aligned sample is indicated by the blue line. For larger single crystals with a thickness of $H$ (sample height), the peak position is shifted as shown by the red line. The misalignment due to the sample height causes a shift of $2x$ along the diffractometer circle [$x = H\cdot{\rm cos}(\theta)$].} \label{sketch}
\end{center}
\end{figure}

The effect of the sample height on the peak position is given by 
\begin{equation}
\Delta 2 \theta_{\rm disp} = S \cdot {\rm cos}(\theta) ~~~~~    {\rm (in~ radians)},
\end{equation}
with
\begin{equation}
S = 2 \cdot \frac{H}{R}
\end{equation}
and $H$ being the sample height and $R$ the diffractometer radius.  
This equation can be derived based on geometrical considerations (Fig.\,\ref{sketch}):
$x$ is given by \(H \cdot {\rm cos}(\theta)\) and with $\Delta 2 \theta_{\rm disp} = 2 \cdot x/R$ (in radians) the above formula follows (the factor of two results from a shift of both incident and outgoing beam). 
For the case of a single crystal the incident beam is depicted by the red line. In fact, here it is not the center of the X-ray beam (dotted line) that fulfills the Bragg equation but the divergent part (straight line). 

The displacement results in a modified Bragg equation: 

\begin{equation}
2 d_{hkl} \cdot {\rm sin}(\theta - \Delta 2\theta_{\rm disp}/2) = \lambda,
\end{equation}

that can be written as

\begin{equation}
2 d_{hkl} \cdot {\rm sin}(\theta - S \cdot {\rm cos}(\theta)/2) = \lambda.
\end{equation}

\begin{figure}
\begin{center}
\includegraphics[width=0.9\textwidth]{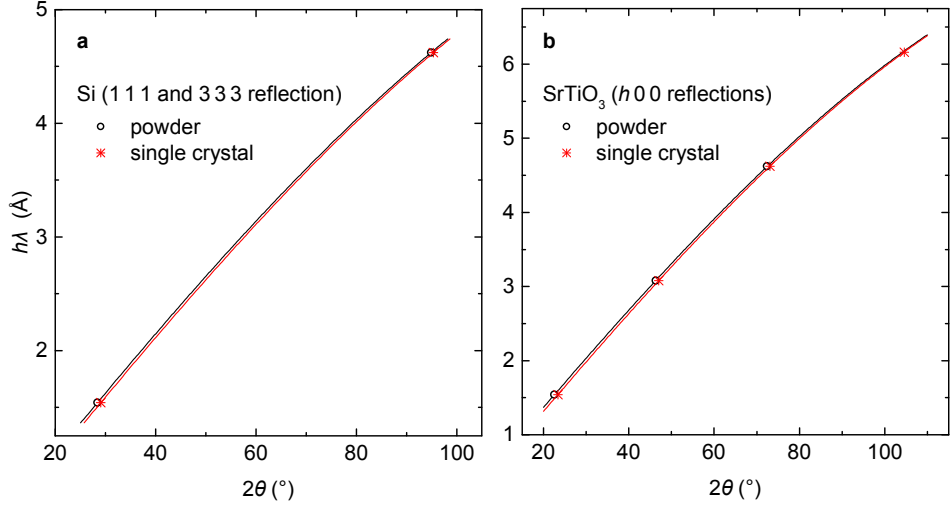}
\caption{Determination of lattice spacing and sample height by fitting $h$ times $\lambda$ as a function of the scattering angle $2\theta$ ($h$ Miller index, $\lambda = 1.5406$\,\AA $\equiv $ Cu-$K\alpha_{1}$ wavelength, see text for details).
a) Silicon powder and single crystal in $(1\,1\,1)$ orientation (wafer). b) Strontium titanate powder and single crystal in $(1\,0\,0)$ orientation. }
\label{fit}
\end{center}
\end{figure}

With two unknowns ($S$ and $d_{hkl}$, the lattice spacing) at least two Bragg reflections are necessary in order to solve the equation. 
Procedures for correcting the displacement have been described earlier, see e.g.\,
\cite[p.363 ff.]{Cullity2001}
Those are based on evaluating the dependence of the apparent lattice spacings on trigonometric functions of $\theta$. 
Given the wide availability of numerical solutions, we propose a different approach: 
Since only one orientation has to be considered, we write the Bragg equation [for ($1\,1\,1$) silicon] as 

\begin{equation}
\frac{2a}{\sqrt{3}} \cdot {\rm sin}(\theta - S \cdot cos(\theta)/2) = \lambda \cdot h,
\label{braggspez}
\end{equation}

with $h = 1$ or 3 for the $1\,1\,1$ and $3\,3\,3$ reflections, respectively ($a$ being the lattice parameter with $d = a/\sqrt{h^2 + k^2 + l^2}$ for a cubic lattice).
It is important to consistently distinguish between radian and degree as unit for the angle; when $\theta$ as the argument of the above function is provided in radians than $S$ is also computed in radians and the sample height follows directly as $H = R \cdot S/2$. 
However, if the full argument of the fitting function is transformed in radians ('inside' the defined fitting function), than S is computed in degree and $H$ has to be multiplied by $2\pi/360$.

In order to allow for a convenient tabular and graphical presentation, we consider the right-hand side of equation\,(5) as dependent variable ("$y$") that is a function of the independent variable $\theta$ ("$x$") and determined by the (unknown) parameters $a$ and $S$.
The fitting was performed with 'Origin 9.1' (Levenberg-Marquardt iteration algorithm), the corresponding plot is shown in Fig.\,\ref{fit}a. 
Such a graphical representation is not necessary in order to calculate the lattice parameters. 
However, it helps to check the fitting results for consistency.
The analogous procedure was performed for strontium titanate (Fig.\,\ref{fit}b), measured using a Rigaku Miniflex 600 diffractometer (Cu-$K\alpha_{1,2}$, Ni-filter, D/teX position sensitive detector). 
The strontium titanate single crystal (a~piece of a wafer) was oriented along ($1\,0\,0$) and 1.0\,mm thick. 
In the given orientation, four reflections are available in the accessible $2\theta$ range. 
The $\sqrt 3$ in the denominator of the modified Bragg equation (eq.\,\ref{braggspez}) has to be dropped since lattice spacing, lattice parameter and Miller index are directly related by $d_{hkl} = a/h$ for the $h\,0\,0$ Bragg peaks of SrTiO$_3$.






\begin{table}
\tbl{Lattice parameter (in \AA) of silicon and strontium titanate single crystals determined with and without correcting for the sample height (in mm) in comparison with corresponding powder measurements (analyzed by LeBail fits) and literature data.}
{\begin{tabular}[l]{@{}lccccc}\toprule
   &  uncorrected 	&  corrected & LeBail fit & literature &sample height  \\
\colrule

Si powder 			&	5.434	&	5.432	& 5.431  & 5.431 $^{\rm a}$ &	$(-)0.06$	\\

Si single crystal		&   5.404	&   5.433	& -      & "            & $(+)0.86$ \\

SrTiO$_3$ powder 		&	3.914	&	3.902	& 3.906 & 3.905 $^{\rm b}$& $(-)0.29$	\\	
	
SrTiO$_3$ single crystal&   3.884	&   3.905 	& - & " & $(+)0.80$ \\

\botrule
\end{tabular}}
\tabnote{$^{\rm a}$Ref.\,\cite{nist640d}}
\tabnote{$^{\rm b}$Ref.\,\cite{Schmidbauer2012}}
\label{tab_si}
\end{table}

The obtained values for lattice parameters and sample height are summarized in Tab.\,\ref{tab_si}. 
Furthermore, the lattice parameters calculated without the correction are given for comparison. 
Those values have been obtained by constraining the sample height to zero, that is the lattice parameter is the only free parameter in the fit of equation\,5 to the $1\,1\,1$ and $3\,3\,3$ peak position whereas $S = 0$. 
Without correcting for the sample height of the Si single crystal, the calculated lattice parameter differs significantly (0.5\%) from the literature value of $a = 5.431$\,\AA \,\cite{nist640d}. 
This value is also obtained as result of a LeBail fit of Si powder (NIST standard) data using the GSAS software package\,\cite{Larson2000, Toby2001}.
The error is even larger (2.1\%) when only the $1\,1\,1$ reflection is considered. 
Similar behavior is observed for strontium titanate: the lattice parameter after correcting for the sample height is in good agreement with the literature value of $a = 3.9053$\,\AA\,\cite{Schmidbauer2012}. 
A LeBail fit of the powder data with GSAS\,\cite{Larson2000, Toby2001} yields $a = 3.906$\,\AA.
The rather large negative sample height observed for the strontium titanate powder is caused by a deeper sample holder.
The sample holder has not been completely filled but a small amount of powder was sprinkled on the bottom in order to resemble a flat sample that is placed with an offset similar to the bottom of the waver. 
In this way, the difference of the sample heights calculated for powder and waver, respectively, should correspond to the thickness of the sample.
A powder measurement of a NIST silicon standard confirmed the good alignment of the diffractometer.

With the modified Bragg equation (4) as given above, a negative sample height corresponds to the sample surface being below the axis of the diffractometer ($\equiv$ theoretical zero-position).  
The small difference between the measured thickness of the Si single crystal (0.78\,mm) and the calculated value of 0.86\,mm - (-0.06\,mm) = 0.92\,mm is caused by a thin layer of grease that has been used for sample mounting and/or by the uncertainty in defining the effective diffractometer radius $R$. 
A similar result was obtained for the strontium titanate single crystal: the difference in the calculated sample heights amounts to 0.80\,mm - (-0.29\,mm) = 1.09\,mm in fair agreement with the measured thickness of 1.0\,mm.

\section{Alignment optimization}\label{tilt}

\begin{figure}
\includegraphics[width=0.98\textwidth]{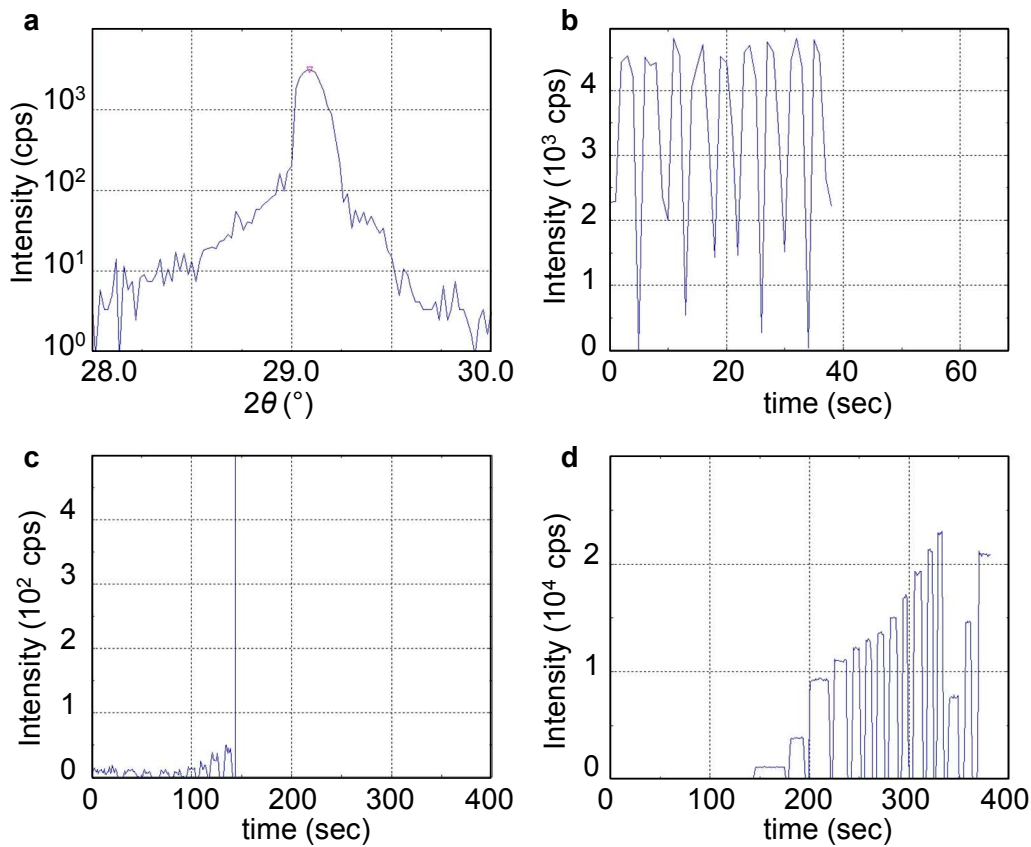}
\caption{Optimization of the sample orientation for higher intensities (demonstrated for a silicon wafer). a) standard $\theta/2\theta$ scan around the $1\,1\,1$ reflection. b) Intensity as a function of time at $2\theta = 29.1^\circ$. The oscillations are caused by the sample-spin that is usually used during measurements of powders in order to improve averaging over crystallite orientations. c,d) Manual rotation of the sample (goniometer) in order to maximize the intensity. An increase in intensity by a factor of 2000 is obtained - note the change of scale for panel d.}
\label{rotate}
\end{figure}

An optimization of the measurement procedure is useful or sometimes necessary if the sample is small and/or misaligned. 
This can be achieved by manually rotating the sample holder such that the tilt of the corresponding (reciprocal) lattice vector points out of the scattering plane. 
Provided the misalignment (tilt) is not too large, the scattered X-Ray beam can still fulfill the Bragg equation and hit the symmetrically positioned detector. 
This holds true since divergence slits are broader in the direction perpendicular to the scattering plane in order to allow for higher intensities. 
An intensional misalignment of a silicon waver by roughly $1^\circ$ had no significant effect on the peak position of the Bragg reflections. 

The proposed alignment can be completed within a few minutes. 
To illustrate the whole procedure with an example, we show the orientation of a silicon wafer step by step.
First, the detector has to be moved to a $2\theta$ position that fulfills the Bragg equation by performing a conventional $\theta/2\theta$ scan in the vicinity of the expected peak position (Fig.\,\ref{rotate}a). 
This is a simple task for the case of the silicon wafer. For misaligned and/or small samples the peak intensity might be low and the peak position not well defined. 
However, a rough estimate within $\pm 0.3^\circ$ is sufficient at this point. For the silicon wafer we find $2\theta = 29.1^\circ$.
Fig.\,\ref{rotate}b shows the intensity at this position as a function of time (the goniometer angle is kept fixed at $\omega = 2\theta/2 = 14.55^\circ$). 
The oscillations are caused by the sample spin that takes place at a frequency of 0.25\,Hz. 
This measurement demonstrates the necessity of the additional alignment: in case of a flat curve it is apparently useless, whereas strong oscillations - as shown here - indicate the possibility to significantly improve the signal-to-noise ratio. A flat curve can be caused by a good alignment and/or a large mosaicity of the single crystal (with lower intensities for the latter). 

In a next step the sample-spin is switched off and the intensity is measured as a function of time. A low intensity of roughly 10 counts per second (cps) is obtained (Fig.\,\ref{rotate}c). 
Next the diffractometer is opened which causes the shutter to close and the intensity to drop to zero instantaneously (depending on the setup, the measurement may have to be interrupted by a software command). 
Now the sample holder is rotated by a few degrees (by hand), the diffractometer closed and the measurement resumed.
At first, no change in intensity is obtained as a result of the performed rotations.  
However, after six repetitions the intensity increases somewhat and the subsequent rotations are performed in smaller steps of roughly one degree. 
A huge increase is observed after step number ten: the intensity increases to $\sim 1000$\,cps (Fig.\,\ref{rotate}d, note that the scale is changed).
A further increase to more than 20,000\,cps is obtained on further careful rotations of the sample holder.  
Eventually, the intensity has been increased by a factor of 2000 compared to the initial values or by roughly one order of magnitude compared with the average of $\sim 3000$\,cps (sample-spin ON, Fig.\,\ref{rotate}a,b). 

Now a regular $\theta/2\theta$ scan is performed with the optimized orientation of the sample holder. 
In case the position of the peak, that had been initially selected for the alignment, changes significantly, the whole procedure should be repeated. 
A measurement performed on a well aligned sample can show a significantly improved signal-to-noise ratio when compared to a conventional measurement (with sample-spin ON). 
This alignment procedure allows for fast measurements with scan rates as high as 10 degrees per minute or more on large enough samples. 
Furthermore, the number of accessible peaks at a given scan speed can be higher. 
Ultimately, samples of a size of $\sim 200\,\mu $m or smaller, which might otherwise require single crystal diffraction techniques, can be measured.

\section{Summary}\label{ende}
When single crystals are measured on a powder diffractometer, the sample height has to be considered or otherwise a significant error is introduced: a displacement of 100\,$\mu$m causes a shift of $\Delta 2\theta = 0.07^\circ$ (for a diffractometer radius of 150\,mm at $2\theta = 20^\circ$). 
In contrast, the zero-shift of a well maintained diffractometer is supposed to be well below 0.05$^\circ$. 
At least two Bragg reflections are necessary to correct for the sample height by calculating the effect on the peak position.
The calculation can be done by a rather simple least-squares fitting in contrast to a sophisticated refinement which requires further efforts and experience. 
When more than two reflections are available, additional corrections can be performed: e.g. for a zero-shift in $2\theta$ or the transparency. For the latter one, a term proportional to sin($2\theta$) could be added in addition to the cos($\theta$) term of the modified Bragg equation. 
However, since the zero-shift highly correlates with the sample displacement and the transparency, these parameters should not be refined together. 

In particular for substitution series the measurement can be restricted to the regions around the expected Bragg reflections (once the phase purity has been settled). 
Even when investigating such substitution series, where the \textit{change} of the lattice parameter with the concentration of a certain component is more important than the absolute values of the lattice parameter, the sample height is important since the aspect ratio (and therefore the average thickness) may also change systematically with the concentration leading to inaccurate results.

To summarize, we describe a fast, accurate, and non-destructive method for measuring the lattice parameters of single crystals using a standard powder diffractometer.

\section*{Acknowledgements}
The authors thank F. Freund, Th. Gr\"unwald, H. Hodovanets, J. Maiwald, S. Ran, and V. Taufour for fruitful discussions. This work was supported by the Deutsche Forschungsgemeinschaft (DFG, German Research Foundation) $-$ JE 748/1 and the U.S. Department of Energy, Office of Basic Energy Science, Division of Materials Sciences and Engineering. The research was performed at the Ames Laboratory. Ames Laboratory is operated for the U.S. Department of Energy by Iowa State University under Contract No. DE-AC02-07CH11358.


\end{document}